\documentclass{JINST}
\let\ifpdf\relax
\usepackage{textcomp}

\title{Testbeam and Laboratory Characterization of CMS 3D Pixel Sensors}

\author{M. Bubna$^{a,m}$\thanks{Corresponding
author.}~, E. Alagoz$^a$, A. Krzywda$^a$,  K. Arndt$^a$, D. Bortoletto$^a$, I. Shipsey$^a$, G. Bolla$^a$, N. Hinton$^a$, A. Kok$^l$, T.-E. Hansen$^l$, A. Summanwar$^l$, J. M. Brom$^j$, M. Boscardin$^b$, J. Chramowicz$^e$, J. Cumalat$^f$, G. F. Dalla~Betta$^c$,  M. Dinardo$^h$,  A. Godshalk$^i$, M. Jones$^a$,  M. D. Krohn$^f$,  A. Kumar$^i$, C. M. Lei$^e$,  R. Mendicino$^c$, L. Moroni$^g$,  L. Perera$^f$, M. Povoli$^c$, A. Prosser$^e$, R. Rivera$^e$, A. Solano$^d$, M. M. Obertino$^d$, S. Kwan$^e$, L. Uplegger$^e$, L. Vigani$^g$, S. Wagner$^f$\\
\llap{$^a$} Physics Department, Purdue University, 525 Northwestern Avenue, West Lafayette, IN 47907-2036, USA\\
\llap{$^b$} Centro per i Materiali e i Microsistemi, Fondazione Bruno Kessler (CMM-FBK), Via Sommarive 18, I-38123 Trento, Italy\\
\llap{$^c$} INFN TIFPA and Dipartimento di Ingegneria Industriale, Universita' di Trento, Via Sommarive 9, I-38123 Povo di Trento (TN), Italy\\
\llap{$^d$} INFN, Sezione de Torino, Via Pietro Giuria, 1, 10125, Turin, Italy\\
\llap{$^e$} Fermi National Accelerator Lab, P. O. Box 500, Batavia, IL 60510-5011, USA\\
\llap{$^f$} University of Colorada at Boulder, Department of Physics, Boulder, CO 80309-0390, USA\\
\llap{$^g$} INFN Milano-Bicocca, Via Celoria 16, I-20126 Milano, 20133 Italy\\
\llap{$^h$} Universit`a and INFN Milano-Bicocca, Piaaza della Scienza 3, I-20126 Milano, Italy\\
\llap{$^i$} State University of New York at Buffalo (SUNY), Department of Physics, Fronczak Hall, Buffalo, NY 14260-1500, USA\\
\llap{$^j$} Institut Pluridisciplinaire Hubert Curien (IPHC)-Inst. Nat. Phys), Strasbourg F-67037 Strasbourg Cedex,
France\\
\llap{$^k$} Department of Physics and Astronomy, 108 Lewis Hall, The University of Mississippi, Mississippi, 38677 (US)\\
\llap{$^l$} Department of Microsystems and Nanotechnology, SINTEF ICT, Gaustadalleen, 0373 Oslo, Norway\\
\llap{$^m$} Department of Electrical and Computer Engineering, Purdue University, 465 Northwestern Avenue, West Lafayette, IN, 47907 USA\\
E-mail: \email{mbubna@purdue.edu}}

\abstract{The pixel detector is the innermost tracking device in CMS, reconstructing interaction vertices and charged particle trajectories. The sensors located in the innermost layers of the pixel detector must be upgraded for the ten-fold increase in luminosity expected at the High-Luminosity LHC (HL-LHC). As a possible replacement for planar sensors, 3D silicon technology is under consideration due to its good performance after high radiation fluence. In this paper, we report on pre- and post- irradiation measurements of CMS 3D pixel sensors with different electrode configurations from different vendors. The effects of irradiation on electrical properties, charge collection efficiency, and position resolution are discussed. Measurements of various test structures for monitoring the fabrication process and studying the bulk and surface properties of silicon sensors, such as MOS capacitors, planar and gate-controlled diodes are also presented.}


\begin{document}

\section{Introduction}\label{sec:Intro}

The current CMS pixel detector ~\cite{bortoletto} was designed to operate up to a fluence of $6\times10^{14}$ $\rm{n_{eq}/ cm^{2}}$, but it is projected to work well up to $1\times10^{15}$ $\rm{n_{eq}/ cm^{2}}$. New radiation hard sensors are required for the High Luminosity upgrade of the LHC (HL-LHC), which is expected to reach an instantaneous luminosity of $L=5\times10^{34}$ $\rm{cm^{-2}s^{-1}}$ and to collect $\approx 3000 $fb$^{-1}$ of data~\cite{LH-LHC}. The innermost layers of pixel detectors will be exposed to a dose of about $10^{16}$ $\rm{n_{eq}/ cm^{2}}$  during the operation of HL-LHC. This is ten times higher than the design fluence of the current detector ~\cite{phase2fluence}. Beam tests data show that  the pixel sensors currently operating in CMS exhibit performance degradation after a fluence of $10^{15}$ $\rm{n_{eq}/ cm^{2}}$~\cite{rohe}, and will not withstand higher radiation doses expected at the HL-LHC. One of the major candidates for replacing the current planar sensor technology is 3D silicon sensors ~\cite{parker}.
In 3D sensors, arrays of p$^+$ and n$^+ $  columns penetrate the bulk. Lateral depletion and smaller electrode spacing yields: (a) shorter carrier drift distance which leads to faster charge collection, (b) lower depletion voltage, (c) smaller trapping probability after irradiation, which leads to superior radiation hardness, and (d) allows implementation of an active edge which reduces the dead region on the edge of the sensor. The smaller inter-electrode spacing in 3D detectors leads to higher capacitance, which increases the sensor noise and degrades the Signal to Noise (S/N) ratio. 3D detectors also require complex processing which is now getting industrialized at various research institutes and companies.	

\begin{figure}[hbt]
\centering
\includegraphics[scale=0.5]{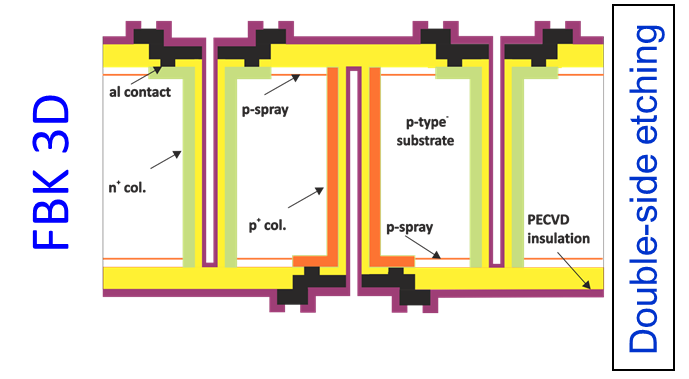}\vspace{-0.2cm} \includegraphics[scale=0.6]{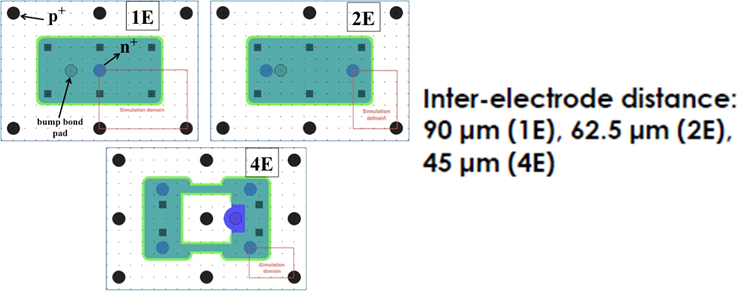}
\caption{Wafer cross-section (left) and pixel layout (right) of FBK 1E, 2E and 4E sensors.}
\label{fig:FBK_layout}
\end{figure}

\begin{figure}[hbt]
\centering
\includegraphics[scale=0.5]{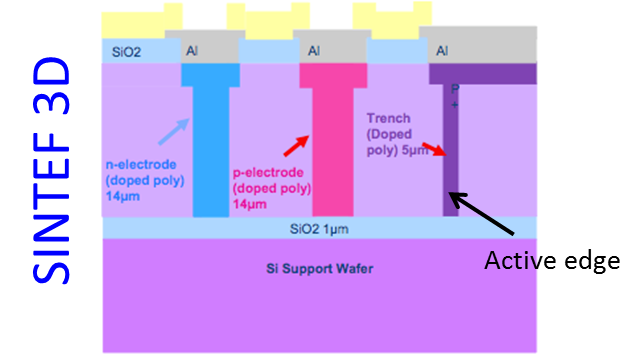}\hspace{0.3cm}\includegraphics[scale=0.5]{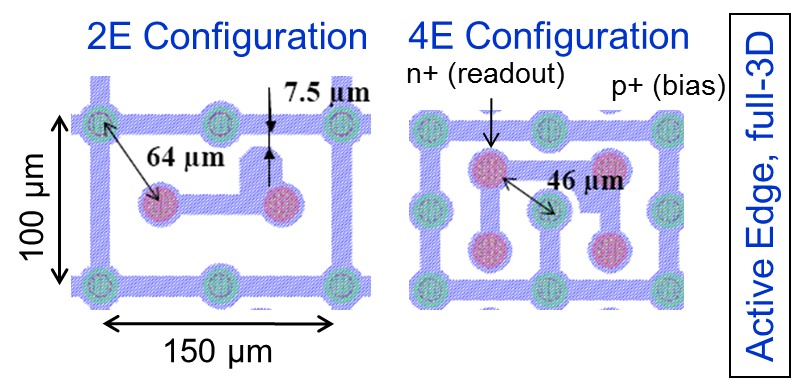}\vspace{0.1cm}
\caption{Wafer cross-section (left) and pixel layout (right) of SINTEF 3D 2E and 4E sensors.}
\label{fig:Sintef_layout}
\end{figure}

\begin{figure}[hbt]
\centering
\includegraphics[scale=0.4,  trim=0 150 20 80]{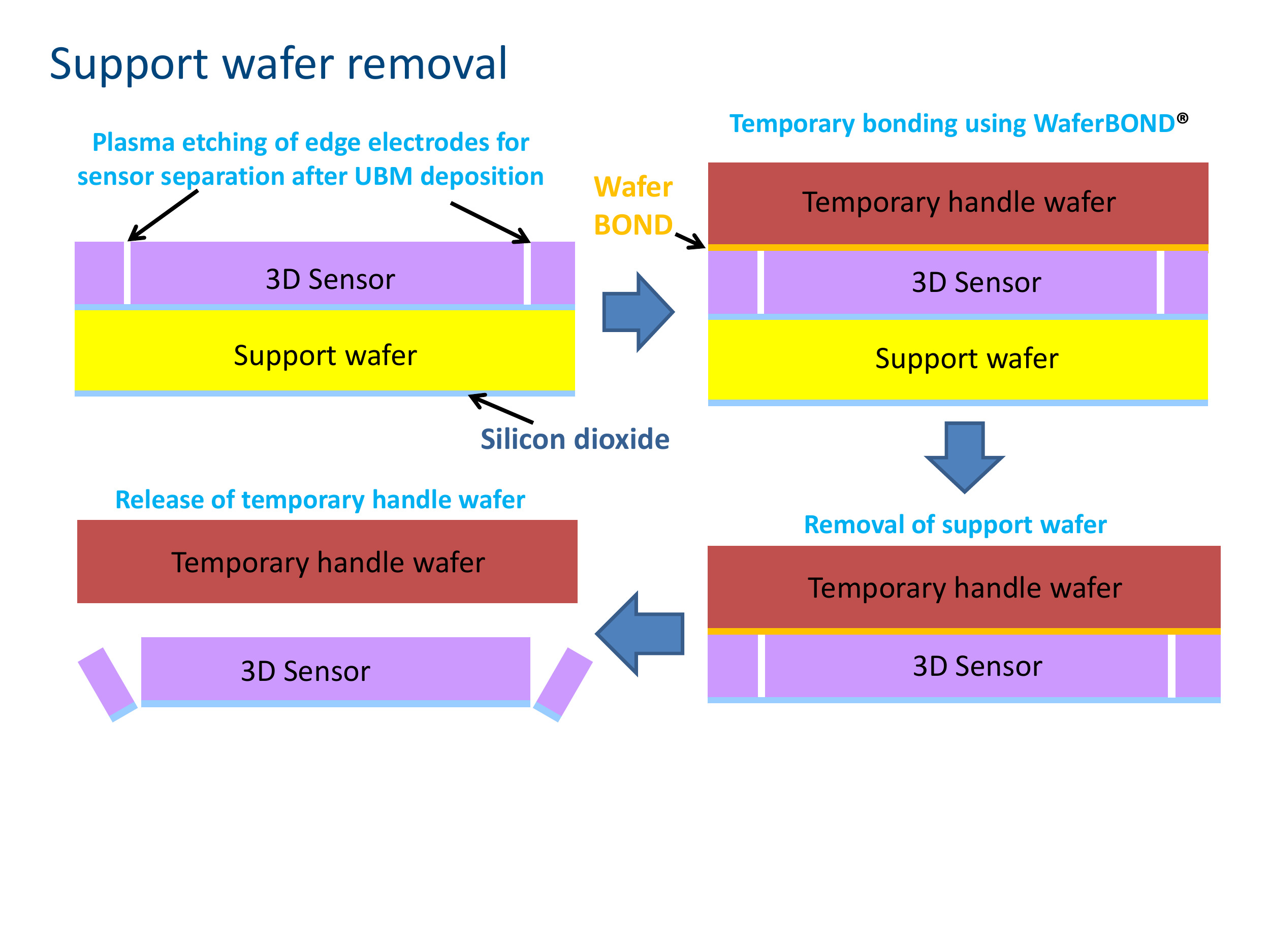}
\caption{Support wafer removal to obtain Active Edge 3D sensors using plasma etching and temporary bonding for supporting the separated sensors.}
\label{fig:Sintef_support_removal}
\end{figure}

In order to reduce the costs and improve yield, Fondazione Bruno Kessler (FBK), Trento, Italy ~\cite{FBK}, and Centro National de Microelectronica (CNM-IMB), Barcelona, Spain ~\cite{CNM}, have independently developed the so-called 3D Double-side Double-Type Column (3D-DDTC) sensors. These designs offer advantages with respect to the original 3D sensor technology, in terms of: (a) reduced process complexity, (b) allowing columnar electrodes without requiring a support wafer, and (c) allowing the back side fully accessible for module assembly. SINTEF, Norway ~\cite{ozhan-nima} has developed the capability of producing full 3D detectors by single-sided processing using a support wafer. By plasma etching instead of dicing, and dopant diffusion of an electrode at the edge of the sensor, an active edge can be obtained which reduces the width of the inactive periphery to only a few microns. Figures ~\ref{fig:FBK_layout} and ~\ref{fig:Sintef_layout} show the cross-section and pixel sensor layout of 3D pixel sensors, manufactured at FBK and SINTEF respectively. One of the challenges of SINTEF technology is the removal of 300 $\mu$m thick support wafer. SINTEF recently removed the support wafer by DRIE (deep reactive ion etching). Temporary bonding using the new WaferBOND$\copyright$ developed by Brewer Science \cite{Brewer} was used to keep the sensors in place once the support wafer was removed. Figure~\ref{fig:Sintef_support_removal} shows the various steps involved in removal of the support wafer from SINTEF 3D wafers.  

In this paper, we considered 3D-DDTC pixel sensors from FBK ATLAS08 (200 $\mu$m thickness), ATLAS11 and ATLAS12 batches (both 230 $\mu$m thickness). A comprehensive overview of FBK 3D technology and the various batches fabricated for the ATLAS Insertable B-Layer production can be found in ~\cite{GG}. We also present results for SINTEF 3D pixel sensors (200 $\mu$m and 285 $\mu$m bulk thickness) which have a trench electrode of \char`\~5 $\mu$m at the edge, making an active edge of \char`\~20 $\mu$m. All the sensors were bump-bonded to the CMS PSI46v2 Read-out Chip (ROC). We also present electrical characterization results for various test structures (Planar diodes, MOS Capacitors) and 1E diodes from FBK ATLAS10 batch. The CMS 1E 3D diode structure from ATLAS10 batch are made from a p-type material with a thickness of 230  $\mu$m and consists of a 19x29 array of 1E pixels shorted together by a metal grid~\cite{povoli}.

The FBK and SINTEF 3D pixel sensors were irradiated at the Los Alamos LANCSE facility with 800 MeV protons to fluences in the range of $7\times10^{14}$ $\rm{n_{eq}/ cm^{2}}$ to $3.5\times10^{15}$ $\rm{n_{eq}/ cm^{2}}$. Also, the 3D diodes from ATLAS10 were irradiated with Co$^{60}$ gamma source to a dose of 2.2 MRad. The 3D diodes were actively biased during gamma ray irradiation. Some of the post-irradiation results for FBK ATLAS08 sensors and ATLAS10 1E diodes were presented earlier in ~\cite{Vienna}.\vspace{-0.2cm}

\section{Electrical characterization}
\subsection{Pre-irradiation IV measurements}
\begin{figure}[hbt]
\centering
\includegraphics[scale=0.4, trim=0 20 0 20]{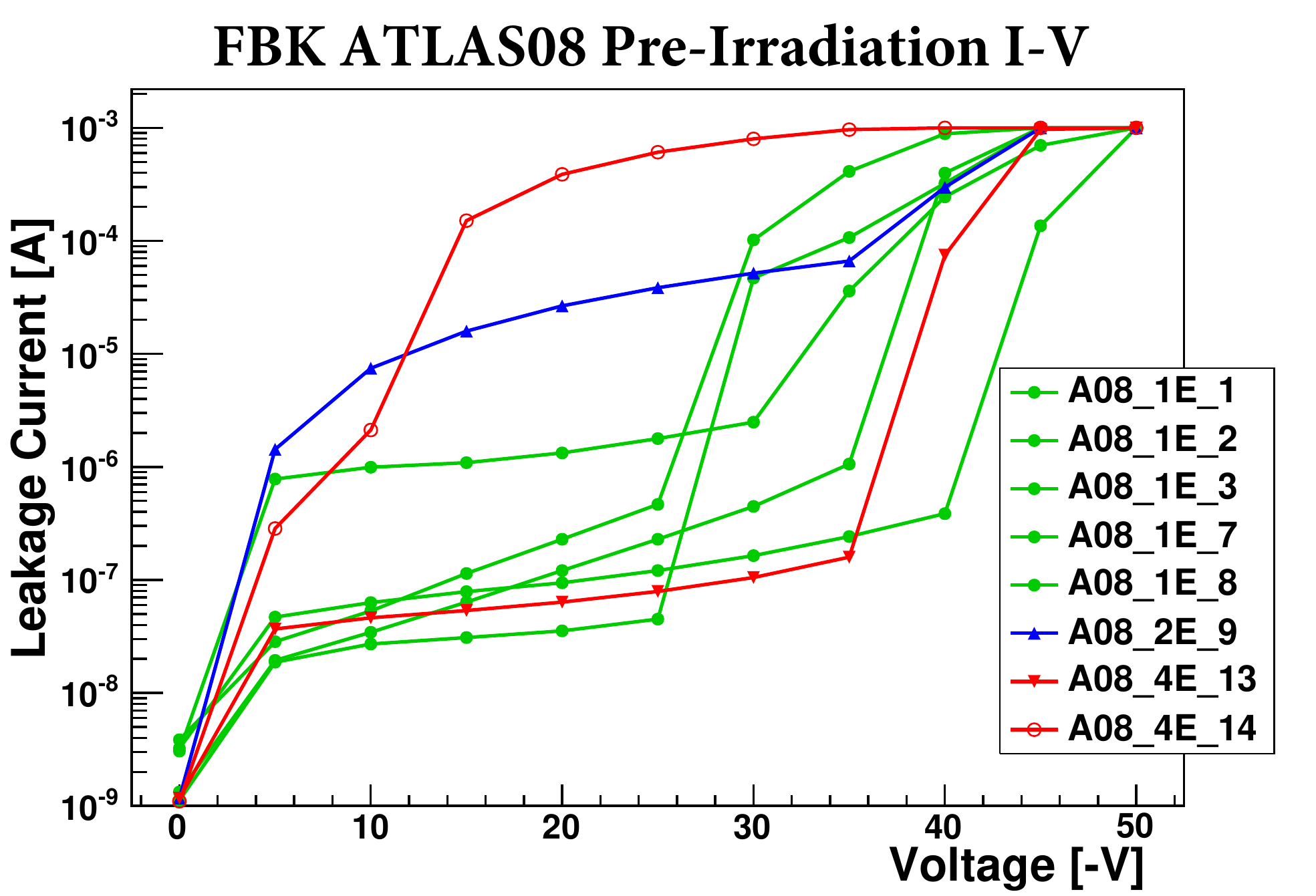}\includegraphics[scale=0.39, trim=0 20 0 20]{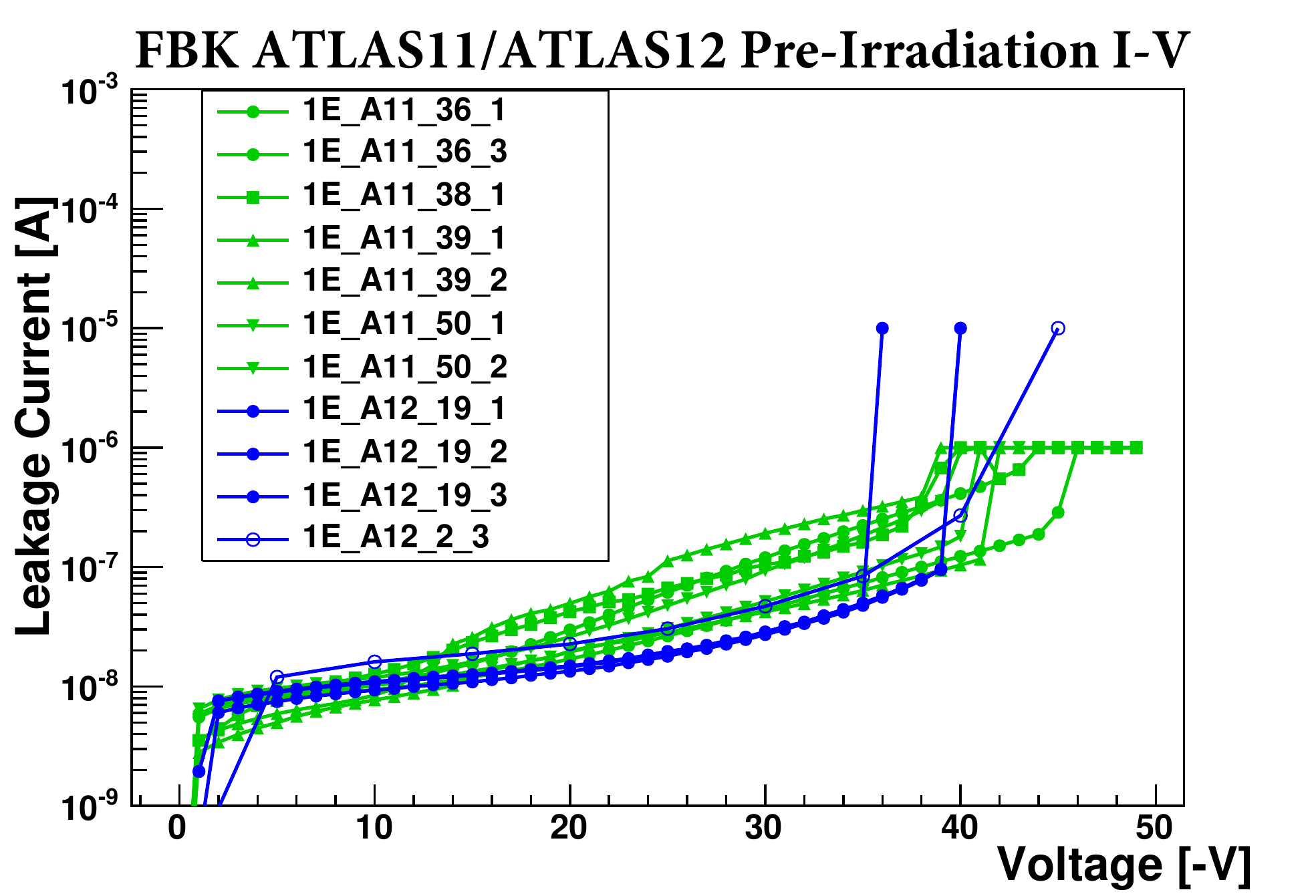}
\caption{I-V measurement of FBK 3D pixel sensors before irradiation at 20$^{\circ}$C from ATLAS08 batch (left) and IBL ATLAS11 and ATLAS12 batches (right).}
\label{fig:FBK_IV}
\end{figure}

The electrical characterization (I-V and C-V measurements) of FBK and SINTEF 3D pixel sensors was performed  in the laboratory at room temperature on-wafer and after dicing. The on-wafer measurements were done using a temporary metal layer which connected all the pixels together, and was later removed as a final processing step. The leakage current, I$_{L}$, was again measured after bump bonding for quality control. The measured I$_{L}$  for assembled modules as a function of the bias voltage, V$_{Bias}$,  for FBK ATLAS08, ATLAS11 and ATLAS12 batch sensors are shown in Figure~\ref{fig:FBK_IV}. ATLAS08 sensors have a depletion voltage of \char`\~5V-10V, I$_{L}$=10 nA -10 $\mu$A, and a breakdown voltage of 35V-45V. ATLAS08 (Figure~\ref{fig:FBK_IV} (left)) was one of the first batches of 3D sensors produced at FBK with fully passing through columns and had some fabrication problems. The inter-pixel isolation in ATLAS08 sensors was done using a p-spray doping of $3\times10^{12}$ $\rm{P/ cm^{2}}$. The p-spray doping caused high electric fields at the junction of the n$^+$ columns and p-spray on the front side of the sensor. The high electric field combined with mechanical stress due to wafer bowing and other surface effects, such as increased oxide charge density and higher surface mobility caused a high leakage current density (1 $\rm{\mu A/cm^{2}}$) and unusually low breakdown voltage in this batch ~\cite{povoli, ealagoz-jinst}. The p-spray doping was reduced to $2\times10^{12}$ $\rm{P/ cm^{2}}$ in ATLAS09 and later batches used for the construction of the ATLAS Insertable B layer (IBL)~\cite{GG}.  As clearly shown in Figure~\ref{fig:FBK_IV} (right), the change in the process  improved the breakdown voltage in IBL sensors, while maintaining sufficient inter-pixel isolation.  FBK ATLAS11 and ATLAS12 1E sensors have better and more uniform I$_{L}$ (10-200 nA) and improved breakdown voltage (35V-45V). The I$_{L}$ of these 3D sensors meets the required specifications and  it is comparable to the current CMS planar sensors (10-100 nA). 
\vspace{-0.0cm}

\begin{figure}[hbt]
\centering
\includegraphics[scale=0.4, trim=140 50 20 0]{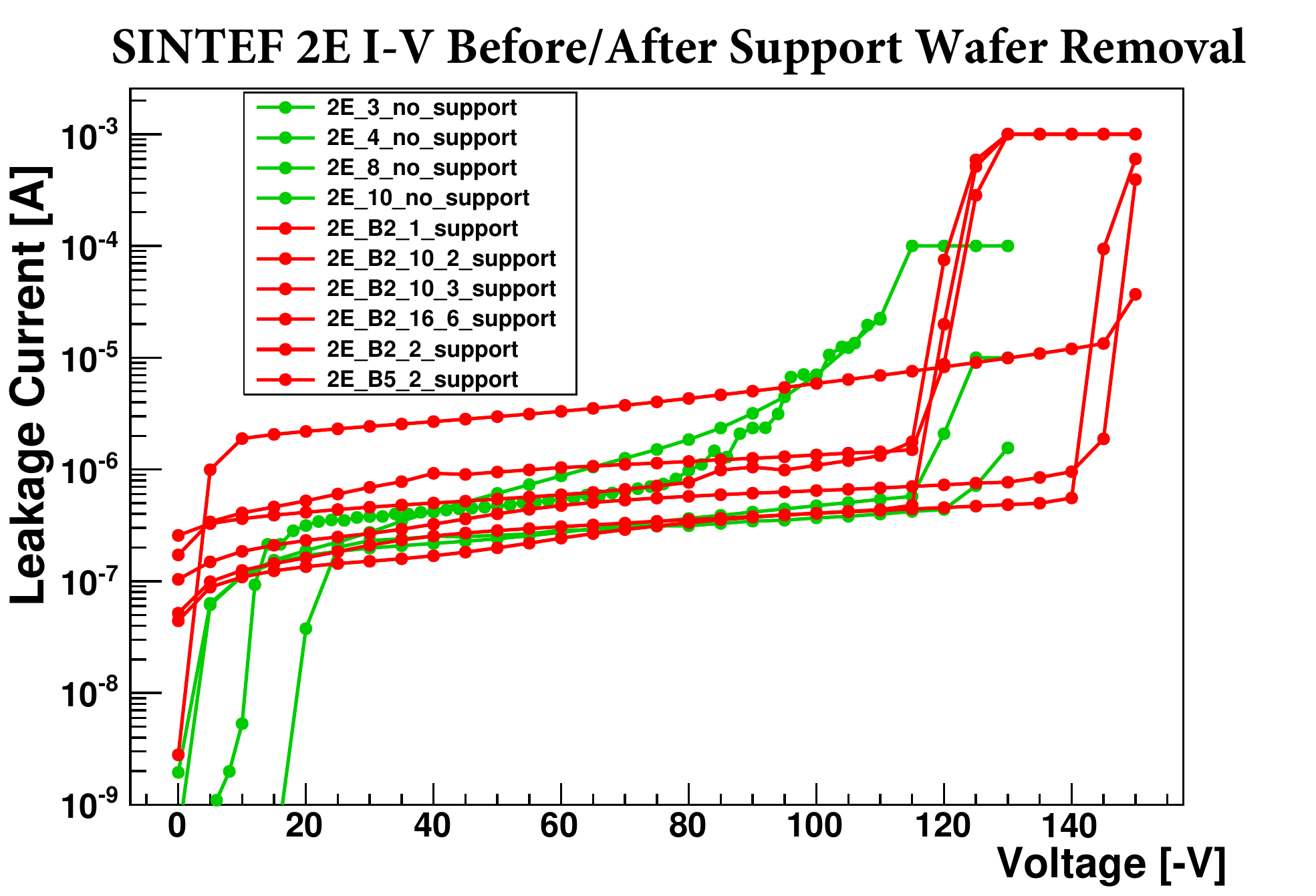}\vspace{0.3cm} 
\includegraphics[scale=0.4, trim=0 50 180 0]{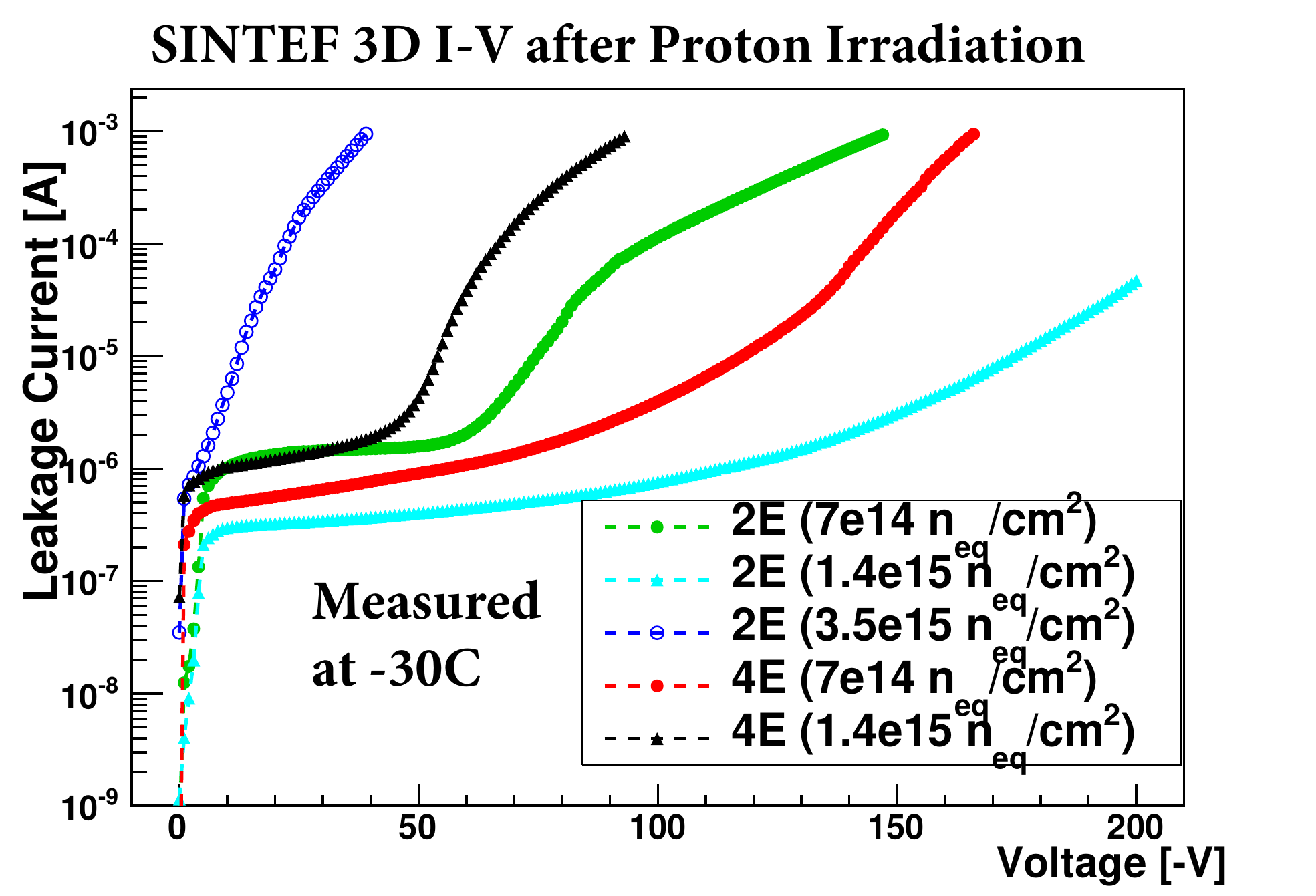}
\caption{I-V measurements of SINTEF 2E sensors before and after support wafer removal (left), Leakage current of SINTEF 3D sensors after $7\times10^{14}$ - $3.5\times10^{15}$ $\rm{n_{eq}/ cm^{2}}$ proton irradiation (right).}
\label{fig:Sintef_IV}
\end{figure}

Similar electrical characterization was done for SINTEF 3D sensors.  The sensors with support wafer were diced and bump bonded to ROCs for measurements and irradiation, while the support wafer was recently removed in some SINTEF sensors. The I-V was measured for assembled 2E and 4E sensors. The measurements show that I$_{L}$ of the 2E sensors (200 nA - 4 $\mu$A) is better than that of 4E sensors (1 $\mu$A - 10 $\mu$A). The sensors get fully depleted at \char`\~10V-20V, the breakdown voltage is 100V-140V for 2E sensors and 90V-110V for 4E sensors. Figure ~\ref{fig:Sintef_IV} (left) shows the leakage current for different SINTEF 2E sensors, before and after the removal of support wafer. The slight decrease in breakdown voltage after support wafer removal for some sensors is due to fabrication defects introduced on the edges of the sensors after separation from the support wafer. These defects are introduced during DRIE (deep reactive ion etching) and other processing steps during support wafer removal, and the corelation between these defects and the increase in leakage current is still under investigation \vspace{-0.2cm}

\vspace{-0.3cm}

\subsection{Post-irradiation IV measurements}
\begin{figure}[hbt]
\centering
\includegraphics[scale=0.37, trim=120 100 50 100]{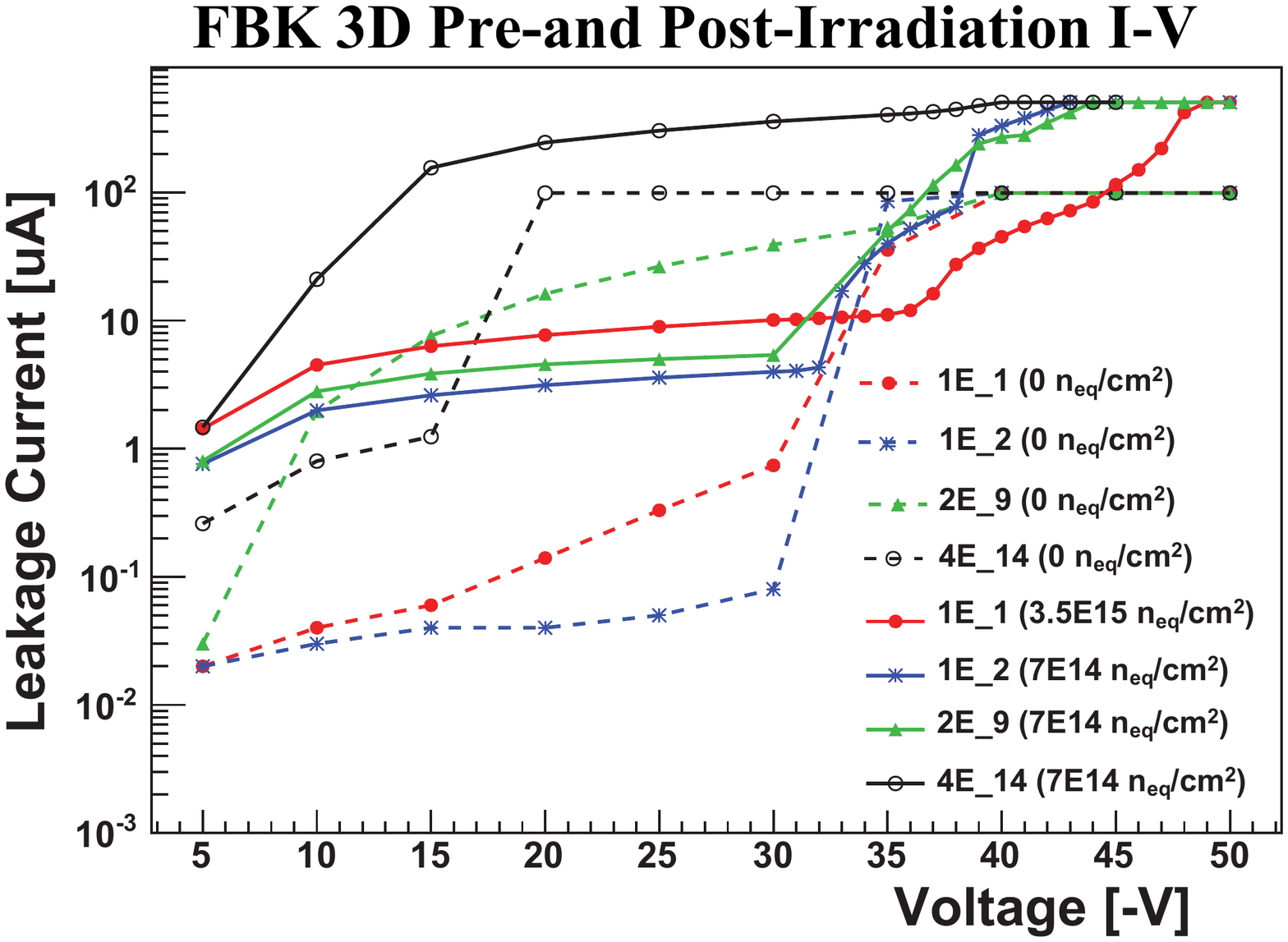}\hspace{0.2cm}\includegraphics[scale=0.41, trim=100 20 50 100]{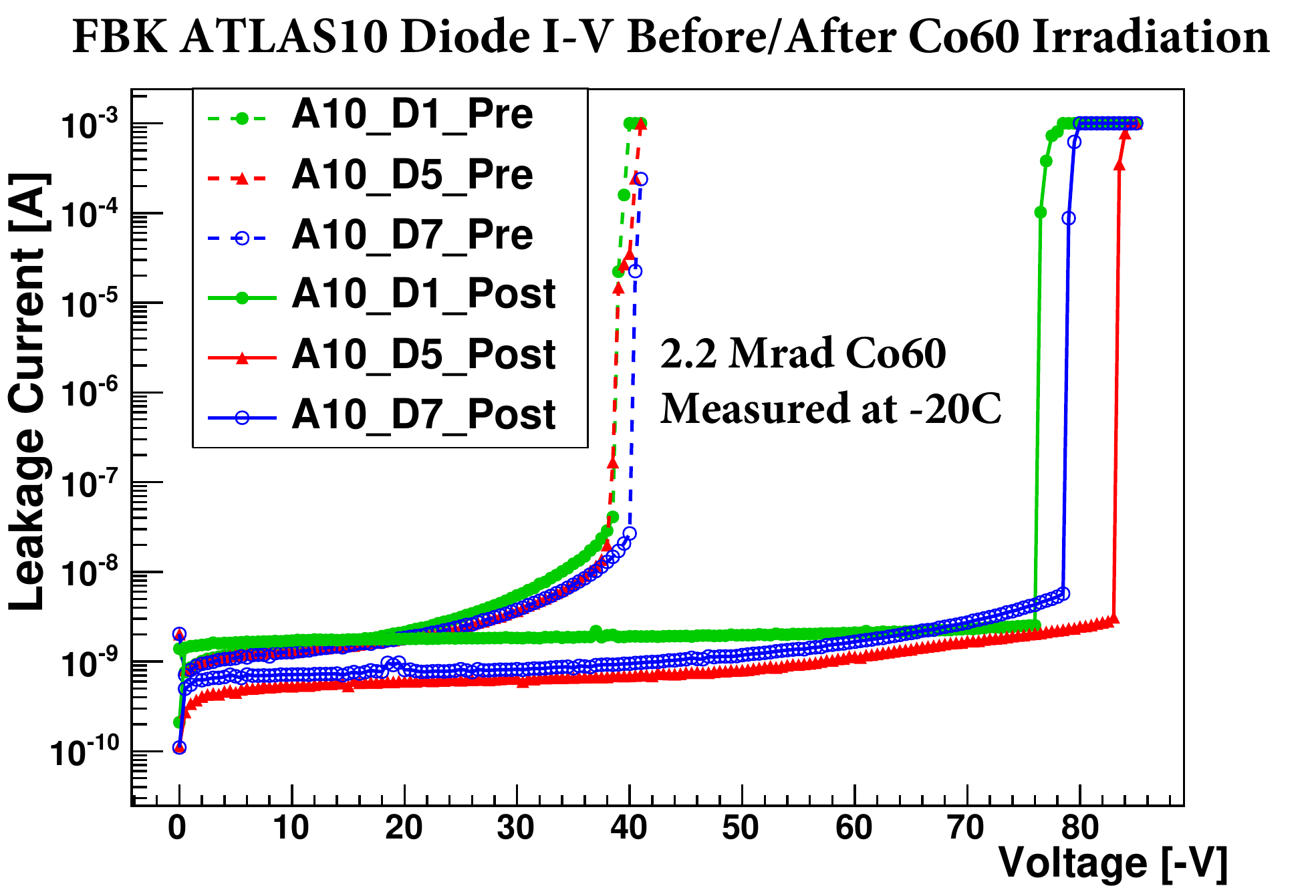}  
\caption{ I-V measurements of FBK 3D ATLAS08 sensors (left) before/after $7\times10^{14}$ - $3.5\times10^{15}$ $\rm{n_{eq}/ cm^{2}}$ proton irradiation, and FBK ATLAS10 1E diodes (right) before/after 2.2 Mrad Co$^{60}$ gamma irradiation.}
\label{fig:FBK_Sintef_Post_IV}
\end{figure}

The leakage current and the breakdown voltage plays an important role in the operation of  the detector, especially after irradiation. In order to fully deplete the sensors, and recover the charge loss after irradiation due to trapping, higher V$_{Bias}$ must be applied to the sensors, which may not always be feasible due to power constraints of the detector. Current CMS High Voltage (HV) supply and cables are supplied by CAEN EASY system and is restricted to 600V ~\cite{CMS-power}. We have studied the surface effects due to irradiation using Co$^{60}$ gamma irradiation and bulk effects with 800 MeV protons.  Figure ~\ref{fig:FBK_Sintef_Post_IV} (left) shows the FBK ATLAS08 sensors measured at -20$^{\circ}$C after $7\times10^{14}$ - $3.5\times10^{15}$ $\rm{n_{eq}/ cm^{2}}$ proton irradiation. The I$_{L}$ in FBK ATLAS08 after proton irradiation increases to 2 $\mu$A - 120 $\mu$A, which is caused by radiation induced defects. The depletion voltage slightly increases to 7V-10V (extracted from CV measurements), while the breakdown voltage slightly improves to 35V-45V. A similar effect was observed in 3D diodes from the same batch (ATLAS08) irradiated with up to 2 Mrad X-rays. This is to be ascribed to the high p-spray implant dose used in ATLAS08 batch, which minimizes the beneficial effect of radiation-induced oxide charge build-up on the breakdown voltage. As already mentioned, in more recent FBK batches ~\cite{GG}, the p-spray dose was reduced, resulting in a higher breakdown voltage before irradiation and in a much larger breakdown voltage increase after irradiation (up to ~160V) ~\cite{IBL}.

This improvement can also be appreciated from Figure ~\ref{fig:FBK_Sintef_Post_IV} (right), which shows the change in the I-V in FBK ATLAS10 diodes after 2.2 Mrad Co$^{60}$ gamma irradiation. As expected, gamma irradiation mostly creates surface defects and there is no trap assisted bulk carrier generation, so the leakage current does not change significantly after irradiation. The breakdown voltage increases from 40V to 80V after irradiation. Gamma irradiation generates a higher concentration of oxide charges that reduces the effective p-spray
doping between the neighbouring pixels. This leads to lower electric fields at the junction of n+ columns and the p-spray region surrounding them and improves the breakdown voltage.

Figure~\ref{fig:Sintef_IV} (right) shows I$_{L}$ for SINTEF 3D sensors after fluences of $7\times10^{14}$ - $3.5\times10^{15}$ $\rm{n_{eq}/ cm^{2}}$  measured at  -30$^{\circ}$C. I$_{L}$ increases to 500 nA - 5 $\mu$A while the breakdown voltage changes to 50V-160V. The I-V distributions show a non-flat slope in many sensors after irradiation, which is due to trap-assisted thermal generation of charge carriers in the band gap region. For the SINTEF 3D sensor which received the dose of $3.5\times10^{15}$ $\rm{n_{eq}/ cm^{2}}$, it was difficult to measure I$_{L}$ due to thermal runaway.

\vspace{-0.3cm}

\section{Laboratory Measurements}

Sensors were tested with a 1mCi Sr$^{90}$ radioactive source in the laboratory. The PSI analog test-board and DAQ software were used as the DAQ system for lab measurements ~\cite{PSI46}. Triggers generated by DAQ system at a fixed frequency of 11 kHz were used to collect data with the source. The hit timestamp in ROC buffer was compared with trigger timestamp generated by the DAQ system to record events with hits.

\subsection{Noise before and after irradiation}
The S-curve test was used to determine the pixel noise by sending internal calibration signals through the injection capacitor to the ROC preamplifier input and measuring the response efficiency on a pixel-by-pixel basis. Noise measurements were taken at room temperature before irradiation, and at $-20^{\circ}$C after irradiation.

\begin{figure}[hbt]
\centering
\includegraphics[scale=0.4, trim=10 20 10 20]{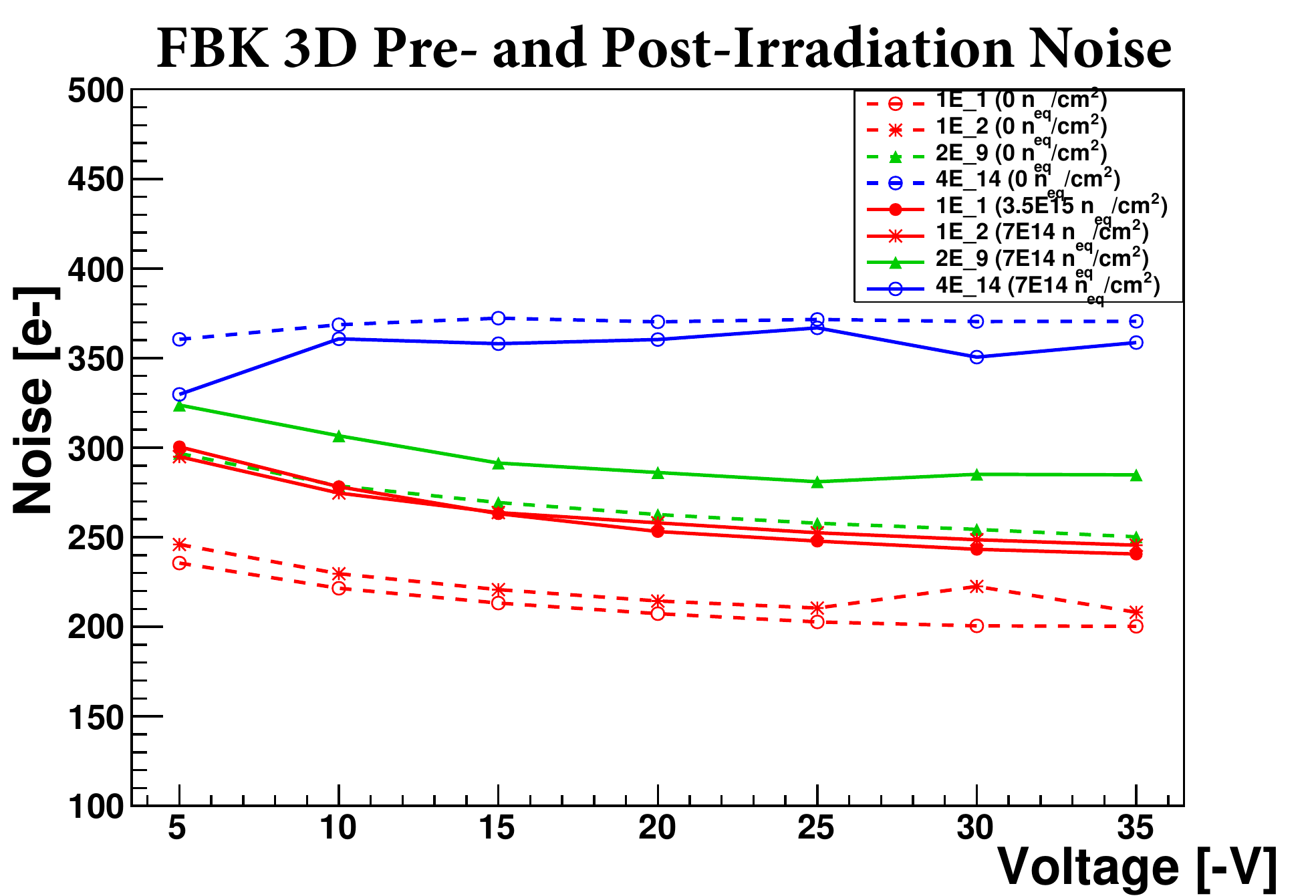}\vspace{0.1cm}\includegraphics[scale=0.4, trim=0 20 0 20]{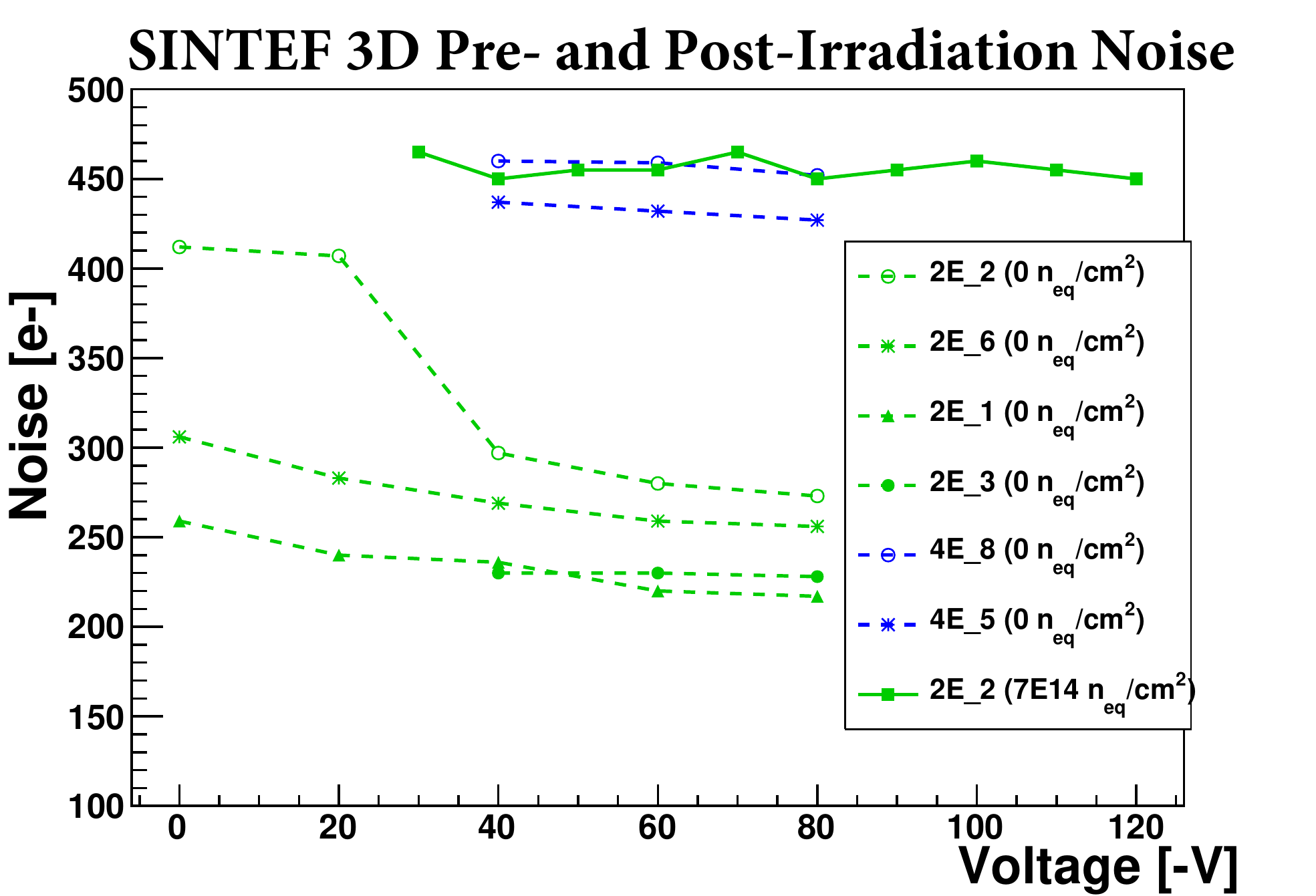}
\caption{Noise before and after $7\times10^{14}$ - $3.5\times10^{15}$ $\rm{n_{eq}/ cm^{2}}$ irradiation: for ATLAS08 sensors (left) and SINTEF 3D sensors (right).}
\label{fig:Noise_FBK_Sintef}
\end{figure}

Figure~\ref{fig:Noise_FBK_Sintef} shows the measured noise before irradiation (dotted) and after irradiation (solid) for FBK ATLAS08 (left) and SINTEF 3D sensors (right). Higher number of electrodes leads to larger capacitance, and higher noise. Before irradiation, the noise of 2E  (4E ) sensors is $\approx$ two (four) times larger than that of 1E and planar sensors. SINTEF 3D sensors have different thickness (200 $\mu$m and 285 $\mu$m bulk with a support wafer of  300 $\mu$m) which explains the difference in noise among various 2E sensors. Capacitance (and thus noise) should not not rise much with radiation. The small increase in noise after irradiation (50$e^{-}$-100$e^{-}$) in FBK sensors is mainly due to increased shot noise caused by increased leakage current. The large increase in noise in SINTEF 2E sensor after irradiation is still under investigation.

\vspace{-0.2cm}

\subsection{Charge collection before and after irradiation}\vspace{-0.1cm}

Charge collection was measured in the laboratory with a Sr$^{90}$ source, before and after irradiation for FBK and SINTEF 3D sensors. The analog ROC was operated with a threshold of 3,900 e$^-$. Figure~\ref{fig:Efficiency_vs_Vthr_Bias} (left) shows the most probable value (MPV) of the charge collected as a function of V$_{Bias}$ before (dotted) and after (solid) irradiation. The amount of charge collected increases at larger values of V$_{Bias}$. Table~\ref{table:Charge} summarises the charge collected in various FBK (200 $\mu$m thickness) and SINTEF (285 $\mu$m thickness) 3D sensors, before and after irradiation. There are small differences between measured charge (as shown in Table~\ref{table:Charge}) and expected values of charge collected (assuming 75 electron-hole pairs generated per micron thickness in silicon at room temperature) for both SINTEF and FBK 3D sensors. There are several possible factors which could explain the lower charge collected in these 3D sensors. One possible reason for the reduced charge collected in SINTEF 3D sensors is the energy loss of beta particles in the support wafer (300 $\mu$m thickness), which the beta particles travel through before reaching the SINTEF sensors. Another possible factor in the mismatch of collected charge in SINTEF 3Ds is due to the reduced charge collection in readout and bias electrodes, which could also explain the secondary peaks at lower charge region in the Landau distribution observed in some SINTEF sensors. There are known variations in wafer thickness for FBK ATLAS08 sensors (200 $\mu$m $\pm$ 20$\mu$m). C-V measurements of planar test diodes for FBK ATLAS08 batch yield a thickness of 185 $\mu$m ~\cite{povoli, ealagoz-jinst} instead of the nominal 200 $\mu$m thickness. In the case of irradiated FBK sensors, early sensor breakdown meant that only relatively lower bias voltage can be applied to the sensor, in particular for the 1E sensors irradiated at the largest fluence. Thus, the sensors were operated partially depleted which reduced the charge collected. Another parameter affecting the charge collection is charge sharing. Due to the 1 cm gap between Sr$^{90}$ source and 3D sensors, electrons arrive at the sensor at a non-zero angle. This can increase the charge sharing between neighbouring pixels causing a fraction of the charge to be below the threshold of readout chip.

Table~\ref{table:Charge} also shows the charge collected in 3D sensors after irradiation. Unfortunately, the CMS pixel readout chip, which was designed to operate only up to a fluence of $6\times10^{14}$$\rm{n_{eq}/ cm^{2}}$,  remained operational in only one out of the six SINTEF modules which were irradiated. The charge collected in irradiated SINTEF 2E (irradiated to the lowest dose of $7\times10^{14}$ $\rm{n_{eq}/ cm^{2}}$) was 12.9k$e^{-}$, indicating a loss of 25\% at 120V. In FBK ATLAS08 sensors, charge loss was 29\% and 17\% in 2E and 4E sensors respectively after $7\times10^{14}$ $\rm{n_{eq}/ cm^{2}}$ at 40V.  The charge loss after irradiation decreases with the increase in the number of $n^+$ readout columnar electrodes in the cell. This is expected as the electric field inside the sensor is highest in a 4E sensor which helps to offset the charge loss due to radiation induced traps. Also, the charge loss  increases as a function of the fluence due to the increased defects generated in the bulk which trap charge carriers. For comparison, the signal loss in CMS planar pixels is about 50\% after  $10^{15}$ $\rm{n_{eq}/ cm^{2}}$ at 600V ~\cite{phase2fluence}, ~\cite{rohe}. Thus, 3D sensors show promising results in the amount of charge collected compared to planar sensors.


By combining the measured charge collected and noise results, the Signal-to-Noise (S/N) ratio for 3D sensors can be calculated. Table~\ref{table:Charge} shows the S/N ratio for various  FBK ATLAS08 and SINTEF 3D sensors, before and after irradiation. The S/N ratio for 1E sensors is highest before irradiation as the 1E sensors have the least value of noise and good charge collection. After irradiation, 2E sensors have the best S/N ratio since they have less charge loss than 1E sensors and lower noise than 4E designs. \vspace{-0.1cm}

\begin{table}
\centering
\begin{tabular}{|c|c|c|c|c|c|}
\hline 
\vtop{\hbox{\strut \textbf{~~~~FBK}}\hbox{\strut \textbf{ATLAS08}}} &\vtop{\hbox{\strut \textbf{Pre-Irradiation}}\hbox{\strut ~~~~~~~\textbf{Charge}}} & \vtop{\hbox{\strut \textbf{Post-Irradiation}}\hbox{\strut ~~~~~~~ \textbf{Charge}}} &  \vtop{\hbox{\strut \textbf{Charge}}\hbox{\strut ~ \textbf{Loss}}} & \vtop{\hbox{\strut \textbf{Pre-Irrad}}\hbox{\strut ~~~~~\textbf{S/N}}} & \vtop{\hbox{\strut \textbf{Post-Irrad}}\hbox{\strut ~~~~~\textbf{S/N}}}\tabularnewline
\hline 
1E & 14.3 k$e^{-}$ &\vtop{\hbox{\strut ~~~~~~~~ 8.2 k$e^{-}$} \hbox{\strut ($7\times10^{14}$ $\rm{n_{eq}/ cm^{2}}$)}}  & 43\% & 66 & 33\tabularnewline
\hline 
1E & 13.8 k$e^{-}$ &\vtop{\hbox{\strut ~~~~~~~~ 6.8 k$e^{-}$} \hbox{\strut ($3.5\times10^{15}$ $\rm{n_{eq}/ cm^{2}}$)}}  & 51\% & 69 & 26.7\tabularnewline
\hline 
2E & 14.8 k$e^{-}$ &\vtop{\hbox{\strut ~~~~~~~~ 10.5 k$e^{-}$} \hbox{\strut ($7\times10^{14}$ $\rm{n_{eq}/ cm^{2}}$)}}  & 29\% & 55 & 38\tabularnewline
\hline 
4E & 14.2 k$e^{-}$ &\vtop{\hbox{\strut ~~~~~~~~ 11.8 k$e^{-}$} \hbox{\strut ($7\times10^{14}$ $\rm{n_{eq}/ cm^{2}}$)}}  & 17\% & 36 & 32\tabularnewline
\hline 
\hline 
\textbf{SINTEF} &\vtop{\hbox{\strut \textbf{Pre-Irradiation}}\hbox{\strut ~~~~~~~\textbf{Charge}}} & \vtop{\hbox{\strut \textbf{Post-Irradiation}}\hbox{\strut ~~~~~~~ \textbf{Charge}}} &  \vtop{\hbox{\strut \textbf{Charge}}\hbox{\strut ~ \textbf{Loss}}} & \vtop{\hbox{\strut \textbf{Pre-Irrad}}\hbox{\strut ~~~~~\textbf{S/N}}} & \vtop{\hbox{\strut \textbf{Post-Irrad}}\hbox{\strut ~~~~~\textbf{S/N}}}\tabularnewline
\hline 
2E & 17.2 k$e^{-}$ &\vtop{\hbox{\strut ~~~~~~~~ 12.9 k$e^{-}$} \hbox{\strut ($7\times10^{14}$ $\rm{n_{eq}/ cm^{2}}$)}}  & 25\% & 60 & 28\tabularnewline
\hline 
4E & 16.9 k$e^{-}$ & - & - & 40 & X\tabularnewline
\hline 
\end{tabular}

\caption{MPV value of charge collected and Signal to Noise (S/N) ratio for FBK and SINTEF 3D sensors before and after proton irradiation.}
\label{table:Charge}
\end{table}
\vspace{-0.1cm}
%
%
%

\section{Testbeam Measurements}
\vspace{-0.2cm}
Sensors were tested with 120 GeV/c protons at the Fermilab Test Beam Facility. The Fermilab testbeam setup is described in ~\cite{lorenzo}. A telescope made of eight planes of planar CMS pixel detectors was used to reconstruct the tracks. The intrinsic track resolution of the telescope is about 7 $\mu$m in both the X and Y local coordinates. The telescope planes are tilted at various angles with respect to the beam to improve resolution. The trigger signal is provided by two PMTs coupled to scintillators, downstream from the telescope.  No magnetic field is applied. Event data from the test beam is analyzed using alignment software developed at INFN Milano, Italy and Fermilab, specifically for the Fermilab beam tests. \vspace{-0.3cm}

\subsection{Pre-irradiation Hit Efficiency}
\begin{figure}[hbt]
\centering
\includegraphics[scale=0.4]{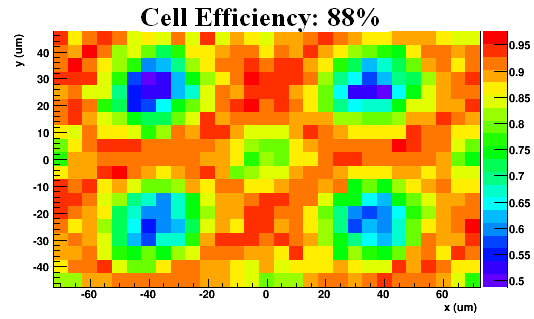}\includegraphics[scale=0.4]{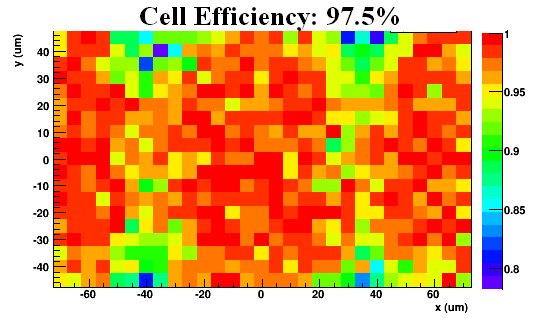}
\caption{Hit efficiency of SINTEF 3D 4E sensor at 0 degree (left) and at 20 degrees (right) measured before irradiation in beam tests.}
\label{fig:Efficiency_Sintef_Pre}
\end{figure}

\vspace{-0.22cm}
The hit efficiency in a pixel cell was measured for 2E and 4E  sensors using events with single tracks. The sensors were operated with a  threshold of 3.9 k$e^{-}$ before irradiation. Figures ~\ref{fig:Efficiency_Sintef_Pre} (left) shows the efficiency for a 4E SINTEF pixel reconstructed from all the tracks information. The sensor was biased at -100V with the detector orthogonal to the beam (angle of $0^{\circ}$). The measured average hit efficiency was $\approx 88$\% for 4E sensor at $0^{\circ}$. The low average hit efficiency is  explained by the charge loss in the bias and readout electrodes which are partially inactive volumes. As shown in Figure~\ref{fig:Sintef_layout}, a SINTEF 4E sensor has four n$^+$ readout columns in the middle of the pixel, and  nine  p$^+$ electrodes corresponding to bias columns surrounding them. The efficiency increases by rotating the Detector Under Test (DUT) at an angle of 20$^\circ$ on the short pitch with respect to the beam axis, reaching a value of 98.7\% for 2E (from 94\% at $0^{\circ}$) and 97.5\% for 4E sensors  (from 88\% at $0^{\circ}$) respectively. This is shown in Figure~\ref{fig:Efficiency_Sintef_Pre} (right) for the 4E sensor. Increasing the angle beyond 20$^\circ$ did not improve efficiency further due to increased charge sharing, which reduced the charge collected inside the pixel below the threshold of the ROC. The loss of efficiency due to the electrodes is less problematic for 1E sensors. 

\begin{figure}[hbt]
\centering
\includegraphics[scale=0.4, trim=20 20 0 20]{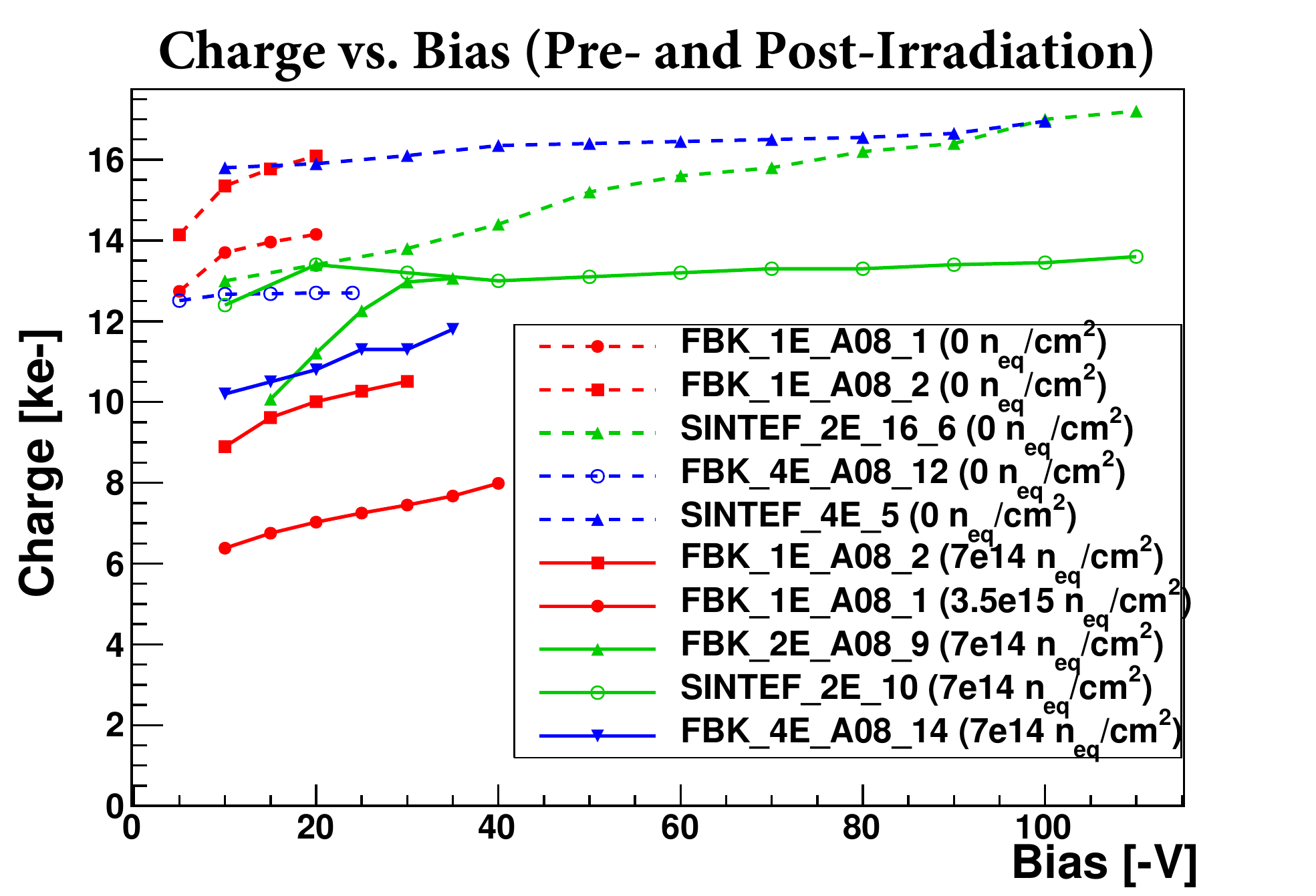}\includegraphics[scale=0.4, trim=20 20 20 20]{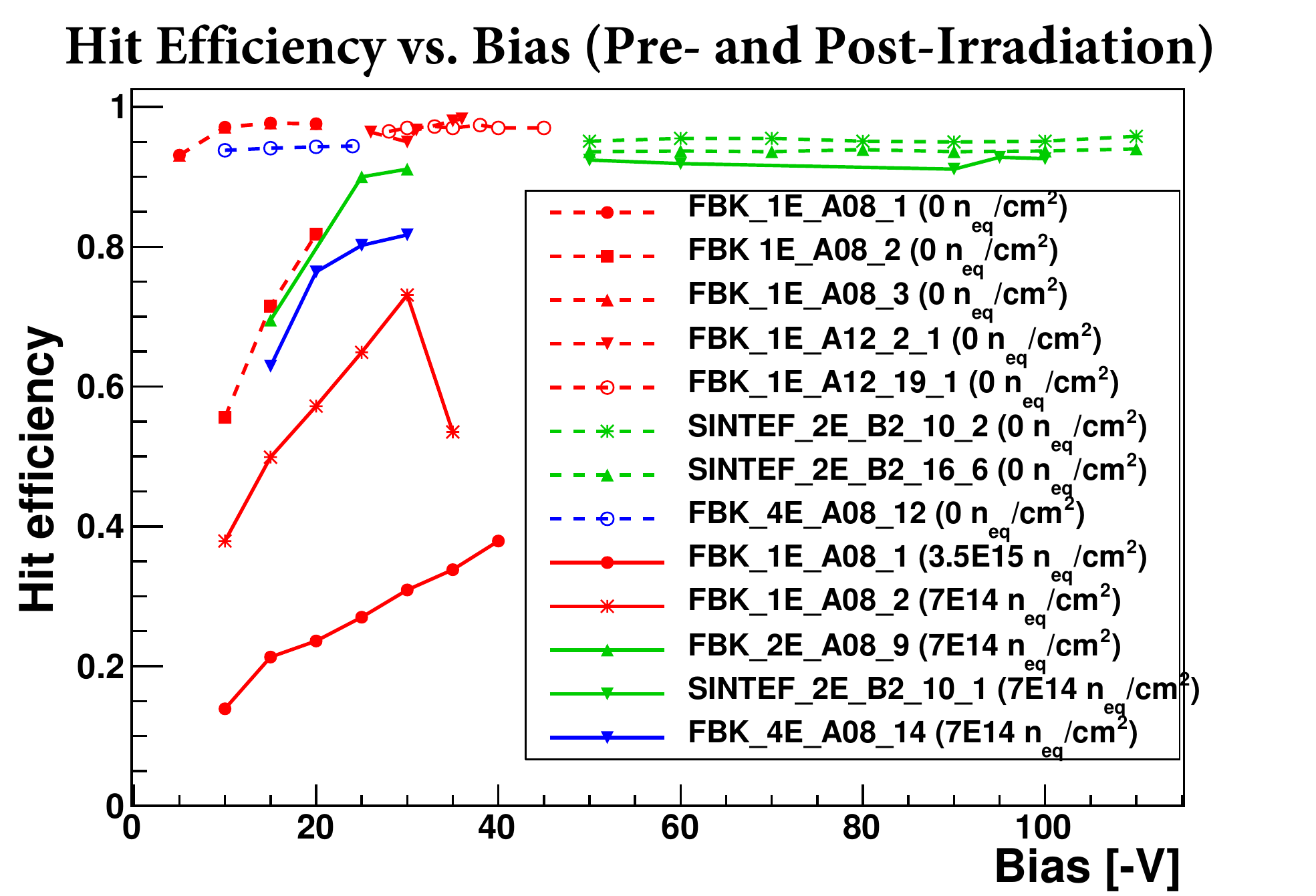}
\caption{Charge vs. Bias using Sr$^{90}$ source (left) and Hit Efficiency vs. Bias measured in beam tests (right) of FBK and SINTEF 3D sensors before and after irradiation.}
\label{fig:Efficiency_vs_Vthr_Bias}
\end{figure}

Figure~\ref{fig:Efficiency_vs_Vthr_Bias} shows the charge collected (left) from Sr$^{90}$ source (discussed earlier in Section 3.2) and the pixel hit efficiency (right) as a function of V$_{bias}$ for the same sensors, before irradiation (dotted curves) and after irradiation (solid curves). The FBK and SINTEF sensors are fully depleted by \char`\~5V and \char`\~10V respectively before irradiation, and the sensors reach their maximum hit efficiency at very low V$_{bias}$. Table ~\ref{table:Cell-efficiency} shows the average value of hit efficiency obtained after $20^{\circ}$ rotation for various FBK and SINTEF 3D sensors. Hit efficiency, even after rotation, decreases going from 1E to 4E layouts since the inactive area due to readout electrodes increases.

\vspace{-0.1cm}

\begin{figure}[hbt]
\centering
\includegraphics[scale=0.4]{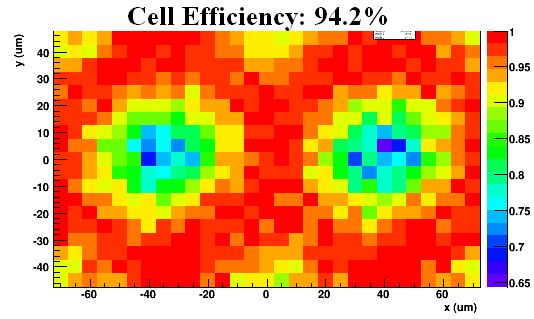}\includegraphics[scale=0.53]{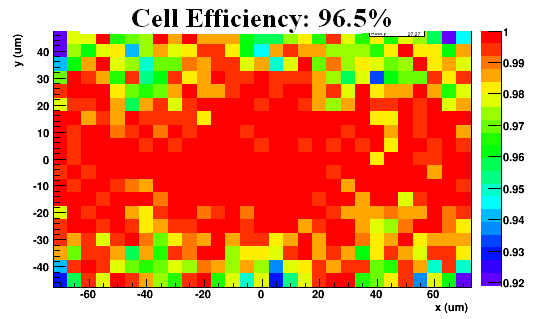}\
\caption{Hit efficiency of SINTEF 2E sensor after $7\times10^{14}$ $\rm{n_{eq}/ cm^{2}}$ proton irradiation: at 0 degrees (left) and at 20 degrees (right)}
\label{fig:Efficiency_Sintef_Post_IV}
\end{figure}
\vspace{-0.1cm}

\subsection{Post-irradiation Hit Efficiency}

After proton irradiation, only one 2E SINTEF module remained operational. The ROC threshold increased by \char`\~2 k$e^{-}$ to 5.9 k$e^{-}$ after proton irradiation for many FBK and SINTEF 3D sensors. This is due to several factors: (a) shift in threshold voltage ($V_t$) and reduced transconductance ($g_m$) of Field-Effect Transistors (FETs) in the readout chip, which slows the chip operation and increases the chip threshold due to time-walk (slower rise time of the pulse height signal), and (b) degradation in sensor performance causing increased noise (shot noise due to leakage current).  Also, after irradiation, the depletion voltage increases, so higher V$_{bias}$ must be applied in order to fully deplete the sensor and recover full charge and hit efficiency. Figure~\ref{fig:Efficiency_Sintef_Post_IV} shows the hit efficiency for a SINTEF 2E pixel sensor after exposure to a fluence of $7\times10^{14}$ $\rm{n_{eq} /cm^{2}}$. The average efficiency of the SINTEF 2E sensor after irradiation is reduced to 94.2\%. By rotating the sensor to 20$^\circ$ with respect to the beam on the short pitch, the average hit efficiency improves to 96.5\% (Figure~\ref{fig:Efficiency_Sintef_Post_IV} (right)). 

Figure~\ref{fig:Efficiency_vs_Vthr_Bias} shows the charge collected (left) and the hit efficiency (right) as a function of the $V_{bias}$. Both quantities typically increase as a function of $V_{bias}$, up to full depletion voltage after which both the charge and efficiency remains constant. However, for many irradiated FBK sensors due to early breakdown after irradiation, the applied bias voltage is not high enough to saturate the charge collected or hit efficiency. Nonetheless the charge collected for all 3D sensors is above 6,000 e$^-$ already at 10V for all 3D detectors types. Table ~\ref{table:Cell-efficiency} summarises  the hit efficiency for various FBK and SINTEF 3D sensors, before and after irradiation. The efficiency loss is a strong function of fluence and rises as the fluence increases. The average hit efficiency  loss of FBK 1E sensors rises from 25\% to  59.9\% after a fluence of $7\times10^{14}$ $\rm{n_{eq}/ cm^{2}}$ and $3.5\times10^{15}$ $\rm{n_{eq}/ cm^{2}}$ respectively. The average hits efficiency loss in FBK ATLAS08 batch and SINTEF 2E sensors ranges between 2-4 \% while FBK 4E sensors  show a loss in average hit efficiency  of 13\% after an irradiation to $7\times10^{14}$ $\rm{n_{eq}/ cm^{2}}$.   2E sensors exhibit the best performance after irradiation. As observed earlier in Section 3.2, the charge loss after irradiation is less in 4E sensors compared to 2E, but the larger inactive volume due to 4 inactive readout electrodes makes the hit efficiency of 4E sensors worse compared to 2E sensors. 

\begin{figure}[hbt]
\centering
\includegraphics[scale=0.4, trim=60 0 0 0]{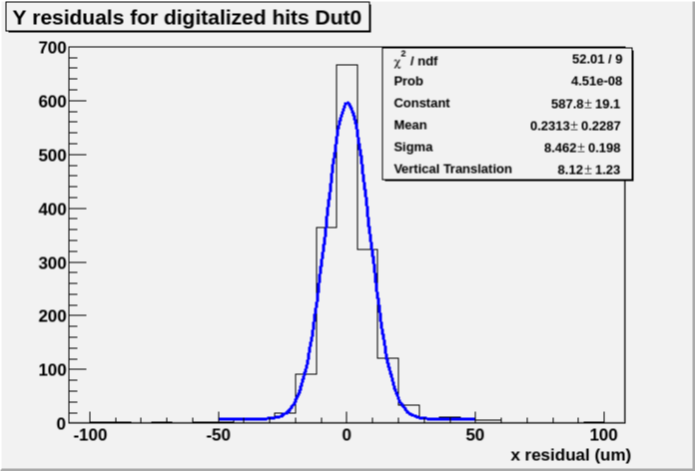}\hspace{0.1cm}\includegraphics[scale=0.4, trim=0 0 60 0]{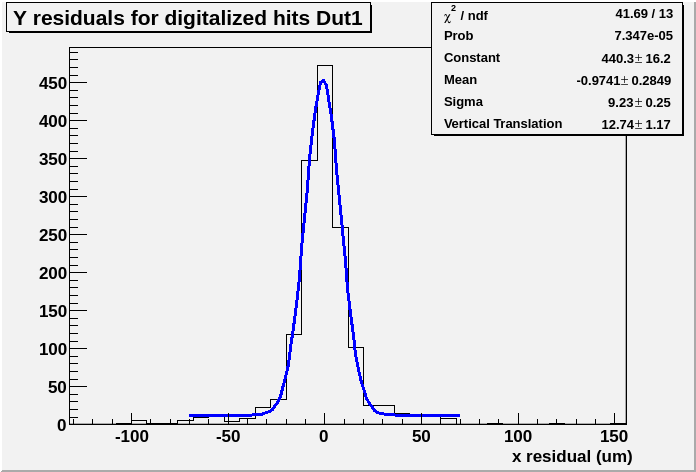}
\caption{Y residual of SINTEF 2E sensor before irradiation (left) and after $7\times10^{14}$ $\rm{n_{eq}/ cm^{2}}$ proton irradiation (right).}
\label{fig:Residual_Sintef_Post_IV}
\end{figure}

Track residuals are calculated as the distance between the predicted and measured positions of a cluster, in either the local X or Y direction. The residuals are fitted with a Gaussian distribution: the overall sensor resolution is determined from the sigma of the Gaussian fit. The uncertainty due to the telescope resolution is
subtracted from the the total track  resolution (including DUTs) to obtain
the DUT resolution. For FBK sensors from ATLAS08 batch, best post-irradiation residual of 12.56 $\mu$m was measured for a 2E sensor. Figure~\ref{fig:Residual_Sintef_Post_IV} shows the measured track residual for a SINTEF 2E sensor, before and after irradiation. The measured track residual was 8.5 $\mu$m before irradiation (left) and 9.2 $\mu$m after irradiation (right).

\begin{table}[hbt]
\centering
\begin{tabular}{|c|c|c|c|}
\hline 
Sensor type &\vtop{\hbox{\strut \textbf{Pre-Irradiation}}\hbox{\strut ~~~~\textbf{Efficiency}}} & \vtop{\hbox{\strut \textbf{Post-Irradiation}}\hbox{\strut ~~~~ \textbf{Efficiency}}} & \textbf{Loss in Efficiency}\tabularnewline
\hline 
FBK ATLAS08 1E\_1 & 97.8\% & 37.9\% (3.5E15  $\rm{n_{eq} /cm^{2}}$)  & 59.9\%\tabularnewline
\hline 
FBK ATLAS08 1E\_2 & 97.6\% & 73.1\% (7E14  $\rm{n_{eq} /cm^{2}}$)  & 25\%\tabularnewline
\hline 
FBK ATLAS08 2E\_9 & 95.4\% & \textbf{91.1\%} (7E14  $\rm{n_{eq} /cm^{2}}$) & \textbf{4\%}\tabularnewline
\hline 
FBK ATLAS08 4E\_12 & 94.5\% & 81.7\% (7E14  $\rm{n_{eq} /cm^{2}}$)  & 13\%\tabularnewline
\hline 
FBK ATLAS12\_1E & 97.3\% & - & -\tabularnewline
\hline 
SINTEF 3D 2E & 98.7\% & \textbf{96.5\%} (7E14  $\rm{n_{eq} /cm^{2}}$) & \textbf{2.2\%}\tabularnewline
\hline 
SINTEF 3D 4E & 97.5\% & - & - \tabularnewline
\hline 
\end{tabular}

\caption{Hit efficiency for various FBK and SINTEF 3D sensors before and after
irradiation.}
\label{table:Cell-efficiency}

\end{table}
\vspace{-0.1cm}

\vspace{-0.2cm}

\section{Measurement of Test Structures}
\begin{figure}[hbt]
\centering
\includegraphics[scale=0.38]{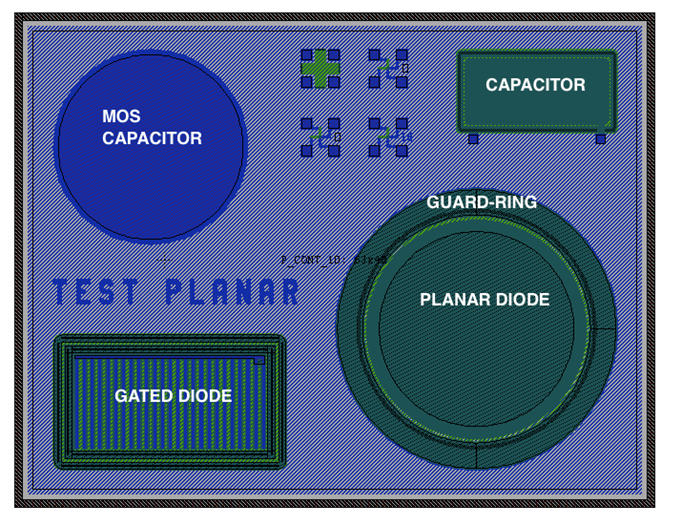}\includegraphics[scale=0.22, trim=100 100 0 0]{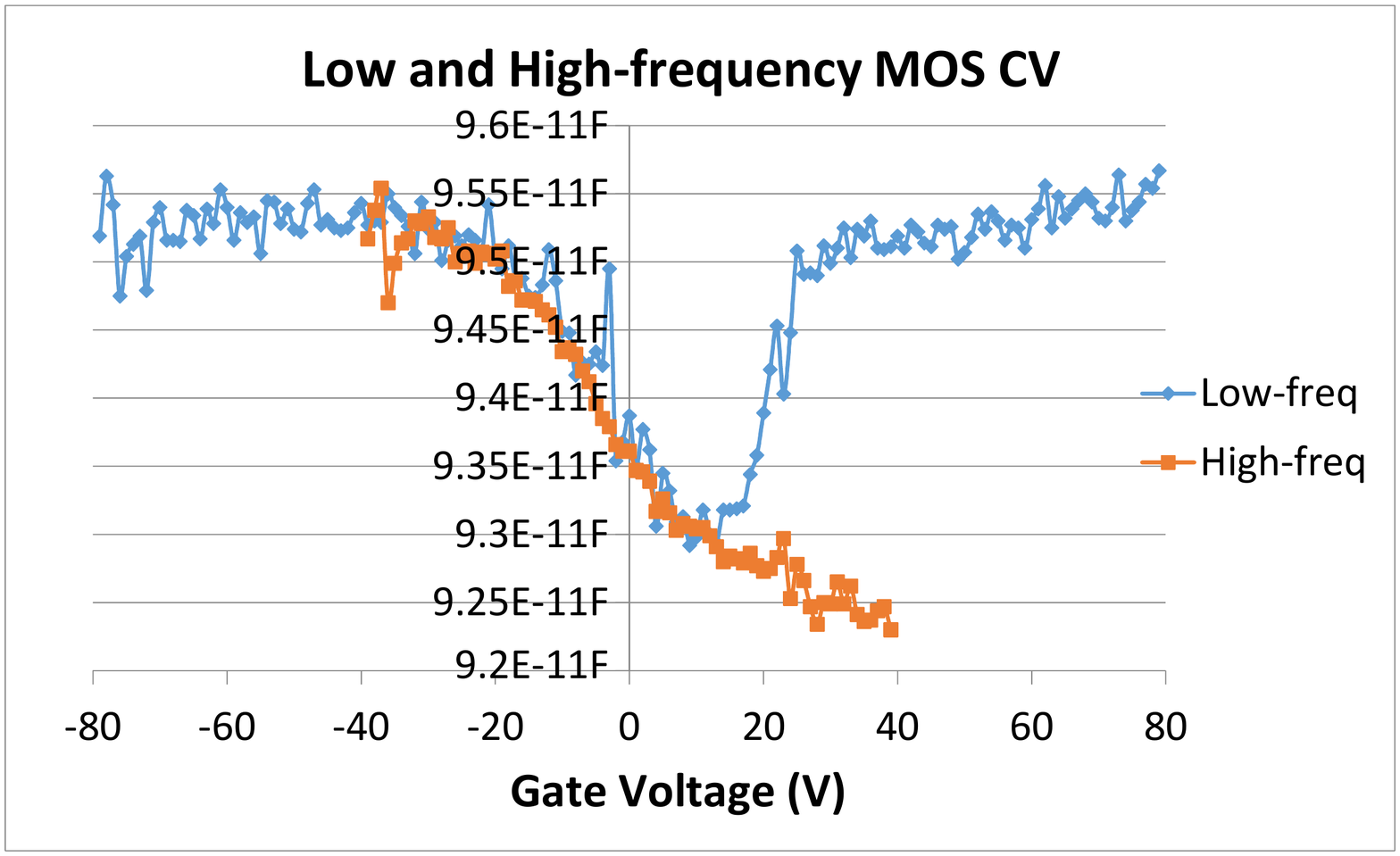}\includegraphics[scale=0.22, trim=100 100 0 0]{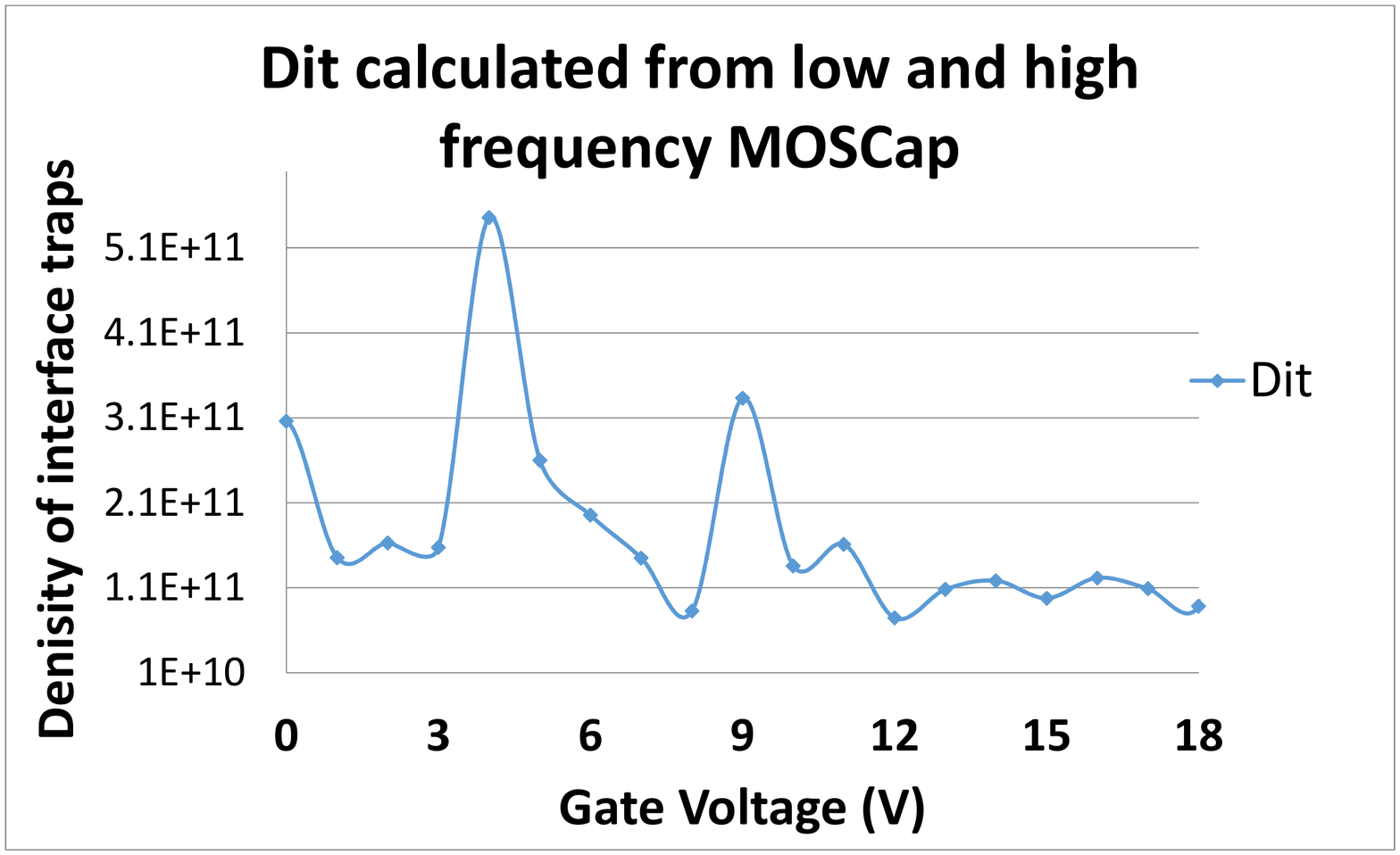}
\caption{Various test structures from FBK ATLAS10 batch (left), low and high frequency CV of MOS capacitor (middle), and measured density of interface traps (right).}
\label{fig:Test_structures}
\end{figure}

The electrical and tracking properties of pixel sensors depend on material properties of bulk silicon and its interface with silicon dioxide. For example, in pixel sensors, interface traps and fixed oxide charge affect the surface current, the breakdown voltage, the capacitance, and the isolation between electrodes. Moreover, ionizing radiation also affects the noise and gain of the front-end readout chip. In order to better understand the bulk and interface properties of silicon sensors, and their effects on sensor properties, various test structures for FBK ATLAS10 and later batches were measured and their material parameters were calculated.  Figure~\ref{fig:Test_structures} (left) shows some of these structures: MOS capacitors, planar diodes, gate controlled diodes and other structures. The measurements include I-V measurements, I-T measurements, and quasi-static and high frequency C-V measurements. Figure~\ref{fig:Test_structures} (middle)  shows the low-frequency (1-2 Hz) and high-frequency (1kHz) C-V measurements measured for MOS capacitors. Figure~\ref{fig:Test_structures} (right) shows the density of interface traps for ATLAS10 sensors calculated by combining quasi-static and high-frequency C-V measurements. The density of interface traps measured was $1.6\times10^{11}$ - $8.9\times10^{11}$ $\rm{traps/ cm^{2}}$. This value is comparable to the ideal value of $10^{8}$-$10^{10}$ $\rm{traps/ cm^{2}}$. Similarly, by measuring I-V and C-V of planar diodes and combining the results, generation lifetime of electrons was measured as $\tau_{g1}=5.78ms$ and $\tau_{g2}=3.64ms$ for two different diodes. Generation and recombination lifetimes help to understand the leakage current and charge collection results.

 \vspace{-0.3cm}

\section{Conclusions and Outlook}\vspace{-0.2cm}
This paper presents measurements of  3D sensor designs from different vendors. 3D sensors offer several improvements over planar sensors, especially at the high luminosities expected at the HL-LHC. 3D sensors show good charge collection results, both before and after irradiation. 3D sensors also have good tracking efficiency and spatial resolution, but suffer from higher noise compared to planar sensors. The reduced efficiency after irradiation  is affected by the high readout threshold of the current ROC. After irradiation, the 2E electrode configuration showed the greatest average hit efficiency and good charge collection. 2E sensors also offer the best S/N ratio and  provide good tracking efficiency and resolution.

More work needs to be done to reach radiation hardness of $10^{16}$ $\rm{n_{eq}/ cm^{2}}$ and to qualify the performance needed for HL-LHC. TCAD simulation efforts are on-going to understand FBK and SINTEF sensor laboratory and beam test results. The optimisation of the pixel size and thickness is ongoing.  Using digital ROCs with lower thresholds will improve the charge collection efficiency and the tracking efficiency after irradiation. Technology improvements also need to be made in order to aggressively reduce the 3D inter-electrode spacing for improved radiation tolerance.

\end{document}